\documentclass[twocolumn]{aastex631}

\graphicspath{{figure/}}
\usepackage{macros}
\usepackage{hyperref}
\usepackage{enumitem}
\usepackage{multirow}

\defcitealias{Wetzel2016}{W16}
\defcitealias{Chan2018}{C18}
\defcitealias{ElBadry2018a}{E18}
\defcitealias{Hopkins2018b}{H18}
\defcitealias{Wheeler2019}{W19}
\defcitealias{GarrisonKimmel2017}{G17}
\defcitealias{GarrisonKimmel2019a}{G19a}
\defcitealias{GarrisonKimmel2019b}{G19b}
\defcitealias{Samuel2020}{S20}
\defcitealias{AnglesAlcazar2017b}{AA17}
\defcitealias{Ma2018a}{M18}
\defcitealias{Ma2019}{M19}
\defcitealias{Ma2020b}{M20}

\newcommand{\ucdavis}{Department of Physics \& Astronomy, University of California, Davis, CA 95616, USA}
\newcommand{\upenn}{Department of Physics \& Astronomy, University of Pennsylvania, 209 S 33rd St, Philadelphia, PA 19104, USA}
\newcommand{\cca}{Center for Computational Astrophysics, Flatiron Institute, 162 5th Ave, New York, NY 10010, USA}
\newcommand{\uconn}{Department of Physics, University of Connecticut, 196 Auditorium Road, U-3046, Storrs, CT 06269-3046, USA}
\newcommand{\uarizona}{Steward Observatory, University of Arizona, 933 N Cherry Ave, Tucson, AZ 85719, USA}
\newcommand{\zurich}{Department of Astrophysics, Universit\"{a}t Z\"{u}rich, Winterthurerstrasse 190, 8057, Zurich, Switzerland}
\newcommand{\hongkong}{Department of Physics, The Chinese University of Hong Kong, Shatin, Hong Kong, China}

\newcommand{\harvard}{Black Hole Initiative at Harvard University, 20 Garden Street, Cambridge, MA 02138, USA}

\newcommand{\calpoly}{Department of Physics \& Astronomy, California State Polytechnic University, Pomona, CA 91768, USA}
\newcommand{\caltech}{
TAPIR, Mailcode 350-17, California Institute of Technology, Pasadena, CA 91125, USA}
\newcommand{\ucmerced}{Department of Physics, University of California, Merced, 5200 N Lake Road, Merced, CA 95343, USA}
\newcommand{\utaustin}{Department of Astronomy, The University of Texas at Austin, 2515 Speedway, Stop C1400, Austin, TX 78712-1205, USA}
\newcommand{\columbia}{Department of Astronomy, Columbia University, New York, NY 10027, USA}
\newcommand{\ucirvine}{Department of Physics \& Astronomy, 4129 Reines Hall, University of California, Irvine, CA 92697, USA}

\newcommand{\northwestern}{Department of Physics \& Astronomy and CIERA, Northwestern University, 1800 Sherman Ave, Evanston, IL 60201, USA}
\newcommand{\ucsd}{Center for Astrophysics \& Space Sciences, University of California, San Diego, 9500 Gilman Drive, La Jolla, CA 92093, USA}
\newcommand{\princeton}{Department of Astrophysical Sciences, Princeton University, Peyton Hall, Princeton, NJ 08544, USA}
\newcommand{\yaleastro}{Department of Astronomy, Yale University, New Haven, CT 06520, USA}
\newcommand{\yalephysics}{Department of Physics, Yale University, New Haven, CT 06511, USA}
\newcommand{\washington}{Department of Astronomy, University of Washington, Seattle, WA 98195, USA}
\newcommand{\carnegie}{Carnegie Observatories, 813 Santa Barbara St, Pasadena, CA 91101, USA}
\newcommand{\wesleyan}{Department of Astronomy, Van Vleck Observatory, Wesleyan University, 96 Foss Hill Drive, Middletown, CT 06459, USA}
\newcommand{\mitkavli}{Department of Physics, Kavli Institute for Astrophysics \& Space Research, MIT, Cambridge, MA 02139, USA}
\newcommand{\pomona}{Department of Physics \& Astronomy, Pomona College, Claremont, CA 91711, USA}
\newcommand{\telaviv}{School of Physics \& Astronomy, Tel Aviv University, Tel Aviv 69978, Israel}
\newcommand{\manchester}{Jodrell Bank Centre for Astrophysics, University of Manchester, Manchester M13 9PL, UK}
\newcommand{\riverside}{Department of Physics \& Astronomy, University of California, Riverside, CA 92521, USA}
\newcommand{\rpi}{Department of Physics, Applied Physics \& Astronomy, Rensselaer Polytechnic Institute, Troy, NY 12180, USA}
\newcommand{\jpm}{J.P.~Morgan Securities LLC, 383 Madison Ave, New York, NY 10017}

\begin{document}

~\vspace{-25 mm}

\title{
\vspace{-8 mm}
Second public data release of the FIRE-2 cosmological zoom-in simulations of galaxy formation}
\shorttitle{FIRE-2 public data release 2}
\shortauthors{Wetzel et al.}

\correspondingauthor{Andrew Wetzel}
\email{awetzel@ucdavis.edu}

\author[0000-0003-0603-8942]{Andrew Wetzel}
\affiliation{\ucdavis}

\author[0000-0002-8429-4100]{Jenna Samuel}
\affiliation{\utaustin}

\author[0000-0003-0965-605X]{Pratik J. Gandhi}
\affiliation{\yaleastro}
\affiliation{\ucdavis}

\author[0000-0002-7484-2695]{Sam B. Ponnada}
\affiliation{\caltech}

\author[0000-0003-1598-0083]{Kung-Yi Su}
\affiliation{\harvard}

\author[0000-0002-8354-7356]{Arpit Arora}
\affiliation{\washington}
\affiliation{\upenn}

\author[0000-0001-5769-4945]{Daniel Angl{\'e}s-Alc{\'a}zar}
\affiliation{\uconn}

\author[0000-0003-4073-3236]{Christopher C. Hayward}
\affiliation{\jpm}

\author[0000-0003-3939-3297]{Robyn E. Sanderson}
\affiliation{\upenn}
\affiliation{\cca}

\author[0000-0002-1109-1919]{Robert Feldmann}
\affiliation{\zurich}

\author{Rachel Cochrane}
\affiliation{\manchester}

\author[0000-0002-3641-4366]{Farnik Nikakhtar}
\affiliation{\yalephysics}

\author[0000-0001-5214-8822]{Nondh Panithanpaisal}
\affil{\carnegie}

\author[0000-0003-1896-0424]{Jose A. Benavides}
\affiliation{\riverside}

\author{Viraj Pandya}
\affiliation{\columbia}

\author{Mike Grudic}
\affiliation{\cca}

\author[0000-0002-3817-8133]{Cameron Hummels}
\affiliation{\caltech}

\author[0000-0002-6145-3674]{Alexander B. Gurvich}
\affiliation{\northwestern}

\author[0000-0001-7326-1736]{Zachary Hafen}
\affiliation{\ucirvine}

\author{Xiangcheng Ma}
\affiliation{\uarizona}

\author{Shea Garrison-Kimmel}
\affiliation{\caltech}

\author{Omid Sameie}
\affiliation{\utaustin}

\author[0000-0003-2544-054X]{T.K Chan}
\affiliation{\hongkong}

\author{Kareem El-Badry}
\affiliation{\caltech}

\author[0000-0003-2806-1414]{Lina Necib}
\affiliation{\mitkavli}

\author[0000-0003-3217-5967]{Sarah Loebman}
\affiliation{\ucmerced}

\author{Sarah Wellons}
\affiliation{\wesleyan}

\author[0000-0002-9497-9963]{Victor H. Robles}
\affiliation{\rpi}

\author{Coral Wheeler}
\affiliation{\calpoly}

\author{Jorge Moreno}
\affiliation{\pomona}

\author{Jonathan Stern}
\affiliation{\telaviv}

\author[0000-0002-9604-343X]{Michael Boylan-Kolchin}
\affiliation{\utaustin}

\author[0000-0003-4298-5082]{James S. Bullock}
\affiliation{\ucirvine}

\author[0000-0002-4900-6628]{Claude-Andr{\'e} Faucher-Gigu{\`e}re}
\affiliation{\northwestern}

\author{Du\v{s}an Kere\v{s}}
\affiliation{\ucsd}

\author{Eliot Quataert}
\affiliation{\princeton}

\author[0000-0003-3729-1684]{Philip F. Hopkins}
\affiliation{\caltech}

\begin{abstract}
We describe the second data release (DR2) of the FIRE-2 cosmological zoom-in simulations of galaxy formation, from the Feedback In Realistic Environments (FIRE) project, available at \href{http://flathub.flatironinstitute.org/fire}{flathub.flatironinstitute.org/fire}.
DR2 includes \textit{all} snapshots for most simulations, starting at $z\!\approx\!99$, with all snapshot time spacings $\lesssim\!25\Myr$.
The \textit{Core} suite---comprising 14 Milky Way-mass galaxies, 5 SMC/LMC-mass galaxies, and 4 lower-mass galaxies---includes 601 snapshots to $z\!=\!0$.
For the \textit{Core} suite, we also release resimulations with physics variations: (1) dark-matter-only versions; (2) a modified ultraviolet background with later reionization at $z\!\approx\!7.8$; (3) magnetohydrodynamics, anisotropic conduction, and viscosity in gas; and (4) a model for cosmic-ray injection, transport, and feedback (assuming a constant diffusion coefficient).
The \textit{Massive Halo} suite now includes 8 massive galaxies with 278 snapshots to $z\!=\!1$.
The \textit{High Redshift} suite includes 34 simulations: in addition to the 22 simulations run to $z\!=\!5$, we now include 12 additional simulations run to $z\!=\!7$ and $z\!=\!9$.
We also release 4 dark-matter-only cosmological boxes used to generate zoom-in initial conditions for many FIRE simulations.
Most simulations include catalogs of (sub)halos and galaxies at all available snapshots, and most \textit{Core} simulations to $z\!=\!0$ include full halo merger trees.
\end{abstract}

\section{Introduction}
\label{sec:intro}

The Feedback In Realistic Environments (FIRE) project\footnote{
\href{https://fire.northwestern.edu}{fire.northwestern.edu}
}
\citep[introduced in][]{Hopkins2014a} seeks to develop cosmological simulations of galaxy formation that resolve the multiphase \ac{ISM}, while implementing the major channels of stellar feedback as directly as possible from stellar evolution models.
FIRE simulations typically zoom in on a region around a single primary galaxy or group of galaxies, to achieve parsec-scale resolution within a cosmological context.

\citet{Wetzel2023} described the first full public data release (DR1) of the FIRE-2 simulations.
DR1 extended our initial data release (DR0) of a subset of FIRE-2 simulations, which comprised 3 \ac{MW}-mass galaxies at $z \! = \! 0$ (m12f, m12i, m12m), accompanied by 9 \textsc{Ananke} synthetic Gaia DR2-like surveys created from these simulations \citep{Sanderson2020}, hosted via \textit{yt Hub} at \href{https://ananke.hub.yt}{ananke.hub.yt}.

This article describes the second public data release (DR2) of the FIRE-2 simulations, which continues to be available at the Flatiron Institute Data Exploration and Comparison Hub (FlatHUB) via \href{http://flathub.flatironinstitute.org/fire}{flathub.flatironinstitute.org/fire}.
DR2 adds to DR1, preserving all contents of DR1.
We briefly describe the simulations and data products that are new in DR2.
See \citet{Wetzel2023} for more comprehensive explanations of the physical models that go into these simulations, contents of snapshots and halo/galaxy catalogs, and analysis tools.

\section{Contents of FIRE-2 Data Release 2}
\label{sec:method}

Table~1 lists all FIRE-2 simulations included in DR2.
DR2 comprises a total of 119 simulations across 4 suites.
Each suite focuses on different mass ranges and different redshifts: the \textit{Core} suite of 20+ simulations to $z \! = \! 0$ (which also includes a subset of simulation run with physics variations), the \textit{Massive Halo} suite of 8 simulations to $z \! = \! 1$, the \textit{High Redshift} suite of 34 simulations to $z \! = \! 5$, and 4 dark-matter-only \textit{Cosmological Boxes} to $z \! = \! 0$.
We describe each of these suites in turn.

\subsection{Core suite to $z \! = \! 0$}
\label{sec:core}

\begin{table*}
\begin{center}
\caption{FIRE-2 simulations included in Data Release 2}
\vspace{-3 mm}
\begin{tabular}{c|c|cccc|c|c}
\hline
\hline
suite & physics & final & \# of & halo mass & \# of & simulation & directory \\
name & variation & redshift & snaps. & [$\Msun$] & sims. & name & \\
\hline
\multirow{21}{*}{Core} & \multirow{6}{*}{Base} & \multirow{6}{*}{0} & \multirow{6}{*}{601} & $10^9$ & 1 & m09 & \multirow{6}{*}{core/} \\
& & & & $10^{10}$ & 2 & m10(q, v) \\
& & & & $10^{11}$ & 6 & m11(b, d, e, h, i, q) \\
& & & & $10^{12}$ & 8 & m12(b, c, f, i, m, r, w, z) \\
& & & & $10^{12} (\times 2)$ & 3 & m12(ThelmaLouise, \\
& & & & & & RomeoJuliet, RomulusRemus) \\
\cline{2-8}
& & \multirow{5}{*}{0} & \multirow{5}{*}{61} & $10^{10}$ & 2 & m10(q, v) & \\
& Dark Matter & & & $10^{11}$ & 5 & m11(d, e, h, i, q) & core/ \\
& Only & & & $10^{12}$ & 9 & m12(b, c, f, i, m, q, r, w, z) & dm\_only/ \\
& & & & $10^{12} (\times 2)$ & 3 & m12(ThelmaLouise, \\
& & & & & & RomeoJuliet, RomulusRemus) \\
\cline{2-8}
& Later & \multirow{2}{*}{0} & \multirow{2}{*}{601} & $10^9$ & 1 & m09 & core/ \\
& Reionization & & & $10^{12}$ & 3 & m12(f, i, m) & reionize\_later/ \\
\cline{2-8}
&\multirow{4}{*}{MHD+} & \multirow{4}{*}{0} & 61 & $10^9$ & 1 & m09 & \\
& & & 61 & $10^{10}$ & 2 & m10(q, v) & core/ \\
& & & 61 & $10^{11}$ & 9 & m11(a, b, c, d, e, h, i, q, v) & mhd/ \\
& & & $61\!-\!309$ & $10^{12}$ & 4 & m12(f, i, m, z) \\
\cline{2-8}
& & \multirow{4}{*}{0} & 61 & $10^9$ & 1 & m09 & \\
& Cosmic & & 61 & $10^{10}$ & 1 & m10v & core/ \\
& Ray & & 61 & $10^{11}$ & 8 & m11(a, b, c, d, e, h, i, v) & cosmic\_ray/ \\
& & & $61\!-\!601$ & $10^{12}$ & 4 & m12(f, i, m, z) & \\
\hline
Massive Halo & Base & 1 & 278 & $0.3\!-\!5\!\times\!10^{13}$ & 8 & A1, A2, A4, A8, B1, B2, C1, C2 & massive\_halo/ \\
\hline
\multirow{8}{*}{High Redshift} & \multirow{8}{*}{Base} & \multirow{4}{*}{5} & \multirow{4}{*}{68} & $10^9$ & 2 & z5m09(a, b) & \multirow{8}{*}{high\_redshift/} \\
& & & & $10^{10}$ & 6 & z5m10(a, b, c, d, e, f) & \\
& & & & $10^{11}$ & 9 & z5m11(a, b, c, d, e, f, g, h, i) & \\
& & & & $10^{12}$ & 5 & z5m12(a, b, c, d, e) & \\
\cline{3-6}
& & \multirow{2}{*}{7} & \multirow{2}{*}{42} & $10^{11}$ & 3 & z7m11(a, b, c) & \\
& & & & $10^{12}$ & 3 & z7m12(a, b, c) & \\
\cline{3-6}
& & \multirow{2}{*}{9} & \multirow{2}{*}{27} & $10^{11}$ & 3 & z9m11(a, b ,c) & \\
& & & & $10^{12}$ & 3 & z9m12(a, b, c) & \\
\hline
Boxes & DM Only & 0 & 11 & - & 4 & L86, L108, L136, L172 & boxes/ \\
\cline{2-6}
\hline 
\hline
\end{tabular}
\end{center}
\vspace{-3 mm}
\tablecomments{
Columns list: 
the name of the suite; the physics variation (``Base'' means base FIRE-2 physics model); the final redshift; the number of available snapshots from $z \! \approx \! 99$ to the final redshift; the approximate halo mass at the final redshift; the number of simulations at that mass; the name of each simulation; the directory at FlatHUB (\href{http://flathub.flatironinstitute.org/fire}{flathub.flatironinstitute.org/fire}).
}
\label{tab:simulations}
\end{table*}

The \textit{Core} suite contains 23 primary galaxies/halos across 20 simulations, comprising 14 \ac{MW}-mass galaxies, 5 SMC/LMC-mass galaxies, and 4 lower-mass galaxies, all run to $z \! = \! 0$.
First, we summarize the additions to the \textit{Core} suite in DR2, which we describe in more detail in the sections below.
DR2 includes:
\begin{itemize}[leftmargin=*]
\item All 601 snapshots for all \textit{Base Physics} simulations from DR1
\item 19 \textit{Dark Matter Only} resimulations
\item 4 \textit{Later Reionization} resimulations that use an updated ultraviolet background, which undergoes cosmic reionization at $z \! \approx \! 7.8$, instead of at $z \! \approx \! 10$ as in other FIRE-2 simulations
\item 16 \textit{MHD+} resimulations that include magnetohydrodynamics, as well as anisotropic conduction and viscosity in gas
\item 14 \textit{Cosmic Ray} resimulations that, in addition to MHD+, also model the injection, transport, and feedback of cosmic rays from supernovae (assuming a constant diffusion coefficient)
\item Catalogs of (sub)halos and galaxies at \text{all} available snapshots, and for most simulations, full merger trees from \textsc{Consistent Trees}
\end{itemize}

\subsubsection{Base Physics}
\label{sec:base}

We refer to the simulations in the \textit{Core} suite that we released in DR1 as having ``Base Physics'', to differentiate them from the additional simulations with physics variations that we release in DR2.
See Table~1 of \citet{Wetzel2023} for the properties of these galaxies at $z \! = \! 0$.
These \textit{Base Physics} simulations are located immediately within the directory \texttt{core/}.
In general, we recommend these as the default galaxy simulations within the \textit{Core} suite to analyze.

For all \textit{Base Physics} simulations, DR2 includes \textit{all} 601 snapshots, starting at $z \! = \! 99$.
Thus, all of these simulations have snapshot time spacing $\lesssim \! 25 \Myr$.
The first snapshot, named $\texttt{snapshot\_000.hdf5}$ or $\texttt{snapdir\_000/}$, corresponds to the same redshift as the initial conditions, which we generated using \textsc{MUSIC} \citep{HahnAbel2011}.
For some simulations, this is $z \! = \! 99$, and for some it is $z \! = \! 100$.

For all \textit{Base Physics} simulations, DR2 includes (sub)halo and galaxy catalogs at all snapshots that have identified halos, and full merger trees across all snapshots.
See Section~\ref{sec:halo} and the Appendix for details.

For the \ac{MW}-mass simulations, we also include low–order basis–function expansion models of the mass distribution of the host halo mass \citep{Arora2022, Arora2024}, provided as spherical-harmonic coefficients for dark matter and hot gas ($T_{\rm gas}\! > \! 10^4$ K) and azimuthal-harmonic coefficients for stars and cold gas, in the directory \texttt{potential/10kpc/}.
Additionally, time-evolving spherical-harmonic expansions of the host-halo dark matter distribution from \citet{Arora2025} are in the directory \texttt{potential/EXP/}.
We also remind users that most \textit{Core} simulations have the exact value of the gravitational potential stored at the location of each particle within each snapshot, specifically, each particle has a stored \texttt{Potential} \citep[see][]{Wetzel2023}.

We request users of the \textit{Core} simulations with \textit{Base Physics} to cite the work that introduced each simulation as listed in Table~1 of \citet{Wetzel2023}.

\subsubsection{Dark Matter Only}
\label{sec:dmo}

In DR2, we release 19 \textit{Dark Matter Only} resimulations, one for almost every simulation in the \textit{Core} suite.
These are located within the subdirectory \texttt{core/dm\_only/}.
Only m09 and m11b do not have a \textit{Dark Matter Only} resimulation.
DR2 also includes a \textit{Dark Matter Only} simulation of an additional \ac{MW}-mass halo, m12q, with $\Mthm \! = \! 1.8 \! \times \! 10^{12} \Msun$ at $z \! = \! 0$ (where $\Mthm$ is the total mass within $\Rthm$, the radius that encloses an average density that is $200 \times$ the mean matter density of the universe), which does not (yet) have a baryonic simulation with FIRE-2 physics at fiducial resolution.

In almost all cases, the resolution in each \textit{Dark Matter Only} simulation matches that of dark-matter particles in the counterpart baryonic simulation.
The exceptions are m10q and m10v, for which the \textit{Dark Matter Only} resimulation has 8 times worse mass resolution (and 2 times larger force softening) than the counterpart baryonic simulation.
For 3 \ac{MW}-mass halos (m12f, m12i, m12m), in addition to a matching-resolution resimulation, we also release a \textit{Dark Matter Only} resimulation with 8 times \textit{better} mass resolution ($m_{\rm dm} \! = \! 5281 \Msun$) and 2 times smaller force softening ($\epsilon_{\rm dm} \! = \! 20 \pc$), which is useful for resolving inner density profiles and low-mass subhalos, including tests of resolution convergence.

For the \textit{Dark Matter Only} simulations, the dark-matter particles contain the same information they do for baryonic simulations, as we listed in \citet{Wetzel2023}.
The key difference is that \textit{Dark Matter Only} simulations do not include gas cells or star particles.
To conserve mass, the dark-matter particles in \textit{Dark Matter Only} simulations are more massive by the cosmic baryon fraction, $m_{\rm dm,dmo} = m_{\rm dm,baryonic} / (1 - \Omega_{\rm b} / \Omega_{\rm m})$, where the values of $\Omega_{\rm b}$ and $\Omega_{\rm m}$ can vary slightly according to the variations in assumed cosmology of each simulation, as the HDF5 header of each snapshot file indicates.
Users should be aware of this when comparing against baryonic simulations and think carefully about what the fairest comparison is.
For example, in comparing the virial mass or radius of a halo, one measures only dark-matter particles in a \textit{Dark Matter Only} simulation, while one should use all particles (dark matter, gas, and stars) in a baryonic simulation.
For massive halos ($\Mthm \! \gtrsim \! 10^{11} \Msun$), this is a reasonably fair comparison.
However, if one is comparing the masses of low-mass (sub)halos, such as those around a \ac{MW}-mass galaxy, then to account for the effects of cosmic reionization, stellar feedback, and ram-pressure stripping that remove most of the baryonic mass from such low-mass (sub)halos, we recommend that users reduce the mass of dark-matter particles in a \textit{Dark Matter Only} simulations by the cosmic baryon fraction, via multipling by $1 - \Omega_{\rm b} / \Omega_{\rm m}$, in comparing to the dark-matter mass in a baryonic simulation.

Each \textit{Dark Matter Only} simulation in DR2 includes a subset of 61 snapshots (every tenth), the same subset as for the \textit{MHD+} and \textit{Cosmic Ray} simulations (see Section~\ref{sec:mhd}).
A few \textit{Dark Matter Only} simulations are missing some snapshots (Romeo\&Juliet, Romulus\&Remus, Thelma\&Louise), while m12z has only a single snapshot at $z \! = \! 0$.
See \texttt{notes.txt} within each simulation directory for details.

That said, we include (sub)halo and galaxy catalogs and merger trees generated across all 600 snapshots (see Section~\ref{sec:halo} and the Appendix).

We request that users of the \textit{Dark Matter Only} simulations cite the work that introduced the corresponding baryonic simulation as listed in Table~1 of \citet{Wetzel2023}.

\subsubsection{Later Reionization}
\label{sec:reionize_later}

In DR2, we release 4 ``Later Reionization'' resimulations from the \textit{Core} suite: one ultra-faint galaxy (m09) and 3 \ac{MW}-mass galaxies (m12f, m12i, m12m), located within the subdirectory \texttt{core/reionize\_later/}.
This version of m09 has 8 times worse mass resolution ($m_{\rm baryon} \! = \! 250 \Msun$) than the fiducial simulation with \textit{Base Physics}, but the \ac{MW}-mass simulations have matching resolution.

For context, all other FIRE-2 simulations use the spatially uniform, meta-galactic ultraviolet background from \citet{FaucherGiguere2009}, which reionizes the universe at $z \! \approx \! 10$, rather than at $z \! \approx \! 7.8$, as recent empirical constraints indicate \citep{FaucherGiguere2020}.
As \citet{Wetzel2023} described, almost all FIRE-2 simulations in DR1 (except m09, m10q, m10v, m11b, Romulus\&Remus) also inadvertently suffer from spurious heating from cosmic rays in neutral gas at temperatures $\lesssim \! 1000$\,K at $z \! \gtrsim \! 10$ (before reionization).
The combination of the too-early reionization model and cosmic-ray heating bug suppresses early star formation in low-mass halos and overheats intergalactic gas at these high redshifts.
For most purposes, such as the properties of massive galaxies at lower redshifts, the high-redshift perturbations from these two issues are likely not significant.
Whether or not this cosmic ray heating bug is significant depends on the application, and users should assess it on a case-by-case basis.

The \textit{Later Reionization} resimulations address both of these issues, by fixing the cosmic-ray heating term and by using a modified version of the ultraviolet background from \citet{FaucherGiguere2009}.
We modified this background at $z \! > \! 6$ only (it is unchanged at $z \! < \! 6$), such that hydrogen reionization occurs at $z \! \approx \! 7.8$, consistent with recent empirical constraints.
Given that this primarily affects star formation in low-mass galaxies, we include a resimulation of an ultra-faint galaxy and 3 \ac{MW}-mass galaxies (with their satellite galaxies).
The primary difference in the \textit{Later Reionization} resimulations is enhanced star formation and stellar masses in galaxies with $\Mstar \! \lesssim \! 10^6 \Msun$ at $z \! = \! 0$ (\citealt{Gandhi2022}, Gandhi et al., in prep.).

For \textit{Later Reionization} resimulations of m12f, m12i, and m12m, all particles at all snapshots also store:
\begin{itemize}[leftmargin=*]
\item \texttt{Acceleration} [$h$ km s$^{-1}$ Gyr$^{-1}$] - 3-D acceleration; multiply by $h$ to convert to km s$^{-1}$ Gyr$^{-1}$
\end{itemize}
which users may find useful for dynamical analysis.

DR2 includes all 601 snapshots for these \textit{Later Reionization} simulations, as well as (sub)halo and galaxy catalogs and full merger trees across all snapshots (see Section~\ref{sec:halo} and the Appendix).

We request users of the \textit{Later Reionization} simulations to cite as follows: for m09 cite \citet{Gandhi2022}; for m12f, m12i, m12m cite this article \citep{Wetzel2025}.

\subsubsection{MHD+ and Cosmic Rays}
\label{sec:mhd}

DR2 includes two related physics variations for most (but not all) galaxies in the \textit{Core} suite.
First, we include 16 \textit{MHD+} resimulations, located within the subdirectory \texttt{core/mhd/}.
In addition to the \textit{Base Physics} of FIRE-2, these include magnetohydrodynamics (MHD) and fully-anisotropic Spitzer-Braginskii conduction and viscosity.
Second, we include 14 \textit{Cosmic Ray} resimulations, located within the subdirectory \texttt{core/cosmic\_ray/}.
In addition to MHD+, these model the injection, transport, and feedback of cosmic rays from supernovae (assuming a constant diffusion coefficient).
\citet{Hopkins2020b} introduced both sets of simulations, see their Table~1 for the properties of these galaxies at $z \! = \! 0$.

For details of the implementation of MHD+ physics in the \textsc{Gizmo} code and FIRE-2 simulations, see \citet{HopkinsRaives2016}, \citet{Hopkins2017}, and \citet{Hopkins2018b}.
Analyses of FIRE-2 simulations showed that the addition of MHD+ physics typically does not significantly change galaxy-wide properties \citep{Su2017, Hopkins2020b}, therefore the galaxy-wide properties of these simulations, as listed in Table~1 of \citet{Hopkins2020b}, are generally similar to those of the \textit{Base Physics} simulations, as listed in Table~1 of \citet{Wetzel2023}.
However, MHD+ can have more significant effects on the small-scale \ac{ISM}, such as the properties of giant molecular clouds \citep{Guszejnov2020}.
\citet{Ponnada2022} showed that the magnetic field strengths in these MHD+ simulations of \ac{MW}-mass galaxies agree with observations of the neutral and the ionized \ac{ISM} in the \ac{MW} and nearby galaxies.

For the \textit{Cosmic Ray} simulations in DR2, \citet{Chan2019} and \citet{Hopkins2020b} describe the model for cosmic-ray injection from supernovae shocks and transport in gas, which includes anisotropic diffusion, streaming, adiabatic, hadronic and Coulomb losses.
These \textit{Cosmic Ray} simulations in DR2 all use a ``single energy bin'' model that represents cosmic rays with a single energy density, roughly the energy of protons near the peak of the cosmic-ray spectrum at $1 \! - \! 10$ GeV.
The primary uncertain parameters are the cosmic-ray diffusion coefficient, $\kappa_{||}$, and streaming velocity.
The \textit{Cosmic Ray} simulations in DR2 model cosmic rays as streaming at the local Alfven speed, with a single fixed diffusion coefficient, $\kappa_{||} \! = \! 3 \! \times \! 10^{29} {\rm cm}^2 {\rm g}^{-1}$ (corresponding to a value of 700 in \textsc{Gizmo}'s code units), which is the ``higher'' coefficient model in \citet{Chan2019} and \citet{Hopkins2020b}.
As these and other works showed \citep{Ji2020, Ji2021, Chan2022}, this value of $\kappa_{||}$ provides the best agreement with $\gamma$-ray observations of nearby galaxies.
As \citet{Ponnada2022} showed, this cosmic-ray model weakens the magnetic field strengths (relative to MHD+ only) at low gas density, near the disk-halo interface, but it largely preserves the magnetic field strengths and their observational agreement at higher densities in the \ac{ISM}.

Furthermore, as the above works showed, this value of $\kappa_{||}$ also produces the most meaningful differences from the simulations with only MHD+.
This is because cosmic rays escape the galaxy and can generate non-thermal pressure in the circumgalactic medium that supports cool dense gas that otherwise would accrete into the galaxy, at least in sufficiently massive halos ($\Mthm \! \gtrsim \! 10^{11} \Msun$) at late times ($z \! \lesssim \! 1 \! - \! 2$), thus reducing star formation and stellar masses in these regimes by factors of $2 - 4$.
By contrast, these effects are minimal on lower-mass galaxies or at higher redshifts.
By comparison, smaller values of $\kappa_{||}$ lead to only minor differences from simulations with only MHD+ in all regimes, because cosmic rays take too long to escape dense star-forming gas and lose their energy to collisional hadronic losses, while larger values of $\kappa_{||}$ let cosmic rays escape too efficiently to have appreciable effects, even in the circumgalactic medium.

We emphasize that the ``correct'' approach to modeling cosmic-ray physics remains highly uncertain \citep[for example,][]{RuszkowskiPfrommer2023}.
Recent implementations of cosmic rays in FIRE simulations \citep{Hopkins2021a, Hopkins2021b, Hopkins2021c, Hopkins2022} explored a range of cosmic-ray transport models, tracking multiple bins of cosmic-ray energy, but we do not include those in DR2.
We release this one constant-diffusion model as a reference, which we have analyzed and benchmarked in several works, but we emphasize that this model does not necessarily make unique or preferred predictions for the effects of cosmic rays.

The \textit{MHD+} and \textit{Cosmic Ray} resimulations in DR2 use the original ultraviolet background from \citet{FaucherGiguere2009}, but they do \textit{not} suffer from the spurious heating from cosmic rays at $z \! \gtrsim \! 10$.

As \citet{Hopkins2020b} described, most of the galaxies in the \textit{MHD+} and \textit{Cosmic Ray} suites are resimulations of galaxies from the \textit{Base Physics} suite, that is, they use the same initial conditions at the same resolution.
We note a few exceptions.
First, in terms of the overall galaxy sample, DR2 includes \textit{MHD+} or \textit{Cosmic Ray} resimulations of only 4 \ac{MW}-mass galaxies (m12f, m12i, m12m, m12z).
Furthermore, two galaxies (m10q and m11q) have \textit{MHD+} versions but not \textit{Cosmic Ray} versions.
In terms of resolution, m09, m10q, and m10v have baryonic particle masses $m_{\rm baryon} \! = \! 250 \Msun$, which is 8 times worse resolution than in their \textit{Base Physics} versions.
Finally, DR2 includes 3 simulations in the \textit{MHD+} and \textit{Cosmic Ray} suites that are not in the \textit{Base Physics} suite:
m11a ($m_{\rm baryon} \! = \! 2100 \Msun$), m11d and m11v ($m_{\rm baryon} \! = \! 7100 \Msun$).
Within the \textit{MHD+} suite, the mass of each primary halo and its galaxy at $z \! = \! 0$ is:
\begin{itemize}[leftmargin=*]
\item m11a: $\Mthm \! = \! 4.6 \! \times \! 10^{10} \Msun$, $M_{\rm star,90} \! = \! 6.2 \! \times\!10^7 \Msun$
\item m11d: $\Mthm\!=\!3.2\!\times \! 10^{11} \Msun$, $M_{\rm star,90} \! = \! 4.9 \! \times \! 10^9 \Msun$
\item m11v: $\Mthm \! = \! 3.5 \! \times \! 10^{11} \Msun$, $M_{\rm star,90} \! = \! 2.5 \! \times \! 10^9 \Msun$
\end{itemize}
where $M_{\rm star,90}$ is the stellar mass within $R_{\rm star,90}$, the radius that encloses 90\% of the stellar mass within $20 \kpc$.

Given the additional physics modeled, both the \textit{MHD+} and \textit{Cosmic Ray} simulations contain additional information for each gas cell in their snapshots:
\begin{itemize}[leftmargin=*]
\item \texttt{MagneticField} [Gauss] - 3-D magnetic field strength
\item \texttt{SoundSpeed} [km s$^{-1}$] - sound speed
\end{itemize}
The \textit{Cosmic Ray} simulations also store:
\begin{itemize}[leftmargin=*]
\item \texttt{CosmicRayEnergy} [$10^{10} \, h^{-1} \Msun \, {\rm km}^2 \, {\rm s}^{-2}$] - energy of cosmic rays; multiply by $10^{10} \, h^{-1}$ to get M$_\odot$ km$^2$ s$^{-2}$
\end{itemize}

For most \textit{MHD+} and \textit{Cosmic Ray} simulations, DR2 contains a subset of 61 snapshots (every tenth).
However, 3 \ac{MW}-mass galaxies have more snapshots available.
Within the \textit{MHD+} suite, the number of snapshots is: 127 for m12i, 253 for m12f, and 309 for m12m.
Furthermore, within the \textit{Cosmic Ray} suite, m12f, m12i, and m12m have all 601 snapshots available.

A few simulations have missing snapshots, within the \textit{MHD+} suite (m10q, m11c, m11q) and within the \textit{Cosmic Ray} suite (m09, m10v, m11b, m11d, m11h, m12m).
See \texttt{notes.txt} within each simulation directory.

For both the \textit{MHD+} and \textit{Cosmic Ray} simulations, m12f, m12i, and m12m (only) have (sub)halo/galaxy catalogs and merger trees at all available snapshots (see Section~\ref{sec:halo} and the Appendix).

We request users of the \textit{MHD+} or \textit{Cosmic Ray} simulations to cite \citet{Hopkins2020b}.

\subsection{Massive Halo suite to $z \! = \! 1$}
\label{sec:massive}

The \textit{Massive Halo} suite is located within the directory \texttt{massive\_halo/}.
In DR1, it comprised 4 massive halos (A1, A2, A4, A8) run to $z \! = \! 1$.
See Table~2 of \citet{Wetzel2023} for the properties of these galaxies at $z \! = \! 1$, and see \citet{Feldmann2016, Feldmann2017, AnglesAlcazar2017b} for more details of these simulations.

DR2 provides two updates.
First, we now include 4 additional halos (B1, B2, C1, C2), which are even more massive but at lower resolution.
\citet{Feldmann2017} introduced these galaxies as run with FIRE-1 physics, and \citet{Cochrane2023a} introduced the versions with FIRE-2 physics that we include in DR2.
B1 and B2 have $m_{\rm baryon} \! = \! 270,000 \Msun$, while C1 and C2 have $m_{\rm baryon} \! = \! 2.2 \! \times \! 10^6 \Msun$.
At $z \! = \! 1$, the mass of each primary halo and its galaxy are:
\begin{itemize}
\item B1: $\Mthm \! \approx \! 1 \! \times \! 10^{13} \Msun$, $\Mstar \! \approx \! 7 \! \times \! 10^{11} \Msun$
\item B2: $\Mthm \! \approx \! 9 \! \times \! 10^{12} \Msun$, $\Mstar \! \approx \! 8 \! \times \! 10^{11} \Msun$
\item C1: $\Mthm \! \approx \!  3 \! \times \! 10^{13} \Msun$, $\Mstar \! \approx \! 2 \! \times \! 10^{12} \Msun$
\item C2: $\Mthm \! \approx \! 5 \! \times \! 10^{13} \Msun$, $\Mstar \! \approx \! 2 \! \times \! 10^{12} \Msun$
\end{itemize}
See \citet{Cochrane2023a} for additional properties of B1, B2, C1, and C2.
Unlike A1, A2, A4, and A8, these new simulations do not track the growth of supermassive black holes.
Because of their lack of AGN feedback, the primary galaxies in all of the \textit{Massive Halo} simulations become overmassive and overdense as they reach $\Mthm \! \gtrsim \! 10^{12} \Msun$ \citep{Wellons2020, Parsotan2021, Cochrane2023b, MercedesFeliz2024}, which is one reason we release snapshots only to $z \! = \! 1$.

Second, for all 8 \textit{Massive Halo} simulations, we now include all 278 snapshots to $z \! = \! 1$, so all snapshot time spacings are $\Delta t \lesssim 25 \Myr$.

For all \textit{Massive Halo} simulations, we include halo/galaxy catalogs generated via the Amiga Halo Finder \citep[\textsc{AHF};][]{Knollmann2009} at (nearly) all snapshots.
For the 4 simulations in DR1 (A1, A2, A4, A8), we continue to include \textsc{Rockstar} \citep{Behroozi2013a} (sub)halo/galaxy catalogs for the subset of 19 snapshots from DR1.
DR2 does not contain halo merger trees for any of the \textit{Massive Halo} simulations.

We request users of the \textit{Massive Halo} simulations to cite \citet{AnglesAlcazar2017b} for A1, A2, A4, A8; and \citet{Cochrane2023a} for B1, B2, C1, C2.

\subsection{High Redshift suite to $z \! = \! 5$}
\label{sec:highz}

The \textit{High Redshift} suite targets halos with $\Mthm \! \approx \! 10^9 \! - \! 10^{12} \Msun$ at $z \! = \! 5 - 9$ \citep{Ma2018a, Ma2018b, Ma2019, Ma2020b}.
See Table~3 of \citet{Wetzel2023} for properties of these galaxies at $z \! = \! 5$.
In addition to the primary galaxy in each simulation, there are many lower-mass (satellite) galaxies within each zoom-in region, so the \textit{High Redshift} suite contains \textit{thousands} of resolved galaxies at each snapshot.

DR2 provides two updates.
First, for the 22 simulations run to $z\!=\!5$ in DR1, DR2 includes all 68 snapshots from $z\!=\!99$ to $z\!=\!5$.
The largest snapshot time spacing (at $z\!=\!5$) is $\Delta t \! < \! 17 \Myr$.
We include \textsc{Rockstar} and \textsc{AHF} halo/galaxies catalogs at (nearly) all snapshots.

Second, we now include the \textit{High Redshift} simulations run to $z\!=\!7$ and $z\!=\!9$, introduced in \citet{Ma2019}, which target more massive galaxies at these higher redshifts than available in the suite to $z\!=\!5$.
Specifically, the suite to $z\!=\!7$ includes 6 galaxies (z7m11a, z7m11b, z7m11c, z7m12a, z7m12b, z7m12c) with 42 snapshots, and the suite to $z\!=\!9$ includes 6 galaxies (z9m11a, z9m11b, z9m11c, z9m11d, z9m11e, z9m12a) with 27 snapshots.
See Table~1 of \citet{Ma2019} for their properties.

For these even higher-redshift simulations, DR2 includes (only) \textsc{AHF} halo/galaxies catalogs at nearly every snapshot.
DR2 does not include halo merger trees for any of the \textit{High Redshift} simulations.

We request users of the \textit{High Redshift} simulations to cite \citealt{Ma2018a}, \citealt{Ma2019}, and/or \citealt{Ma2020b}, as appropriate for the simulations used.

\subsection{Cosmological Boxes to $z \! = \! 0$}
\label{sec:box}

DR2 also includes a suite of 4 uniform-resolution dark-matter-only cosmological boxes that we used to generate the zoom-in initial conditions for many (but not all) FIRE simulations.

We generated most of the initial conditions for earlier FIRE-1 and FIRE-2 simulations (including m10q, m10v, m11b, m11q, m11v, m12b, m12c, m12f, m12i, m12m) using a cosmological box from the AGORA project \citep{Kim2014}.
However, for most of our newer initial conditions, we used the 4 cosmological boxes we release here.
All 4 boxes use the same cosmology, consistent with \citet{Planck2020b}: $\Omega_{\rm m} \! = \! 0.31, \Omega_{\rm \Lambda} \! = \! 0.69, \Omega_{\rm b} \! = \! 0.048, h \! = \! 0.68, \sigma_{\rm 8} \! = \! 0.82, n_{\rm s} \! = \! 0.97, w \! = \! -1$.

These 4 simulations have names that indicate the comoving side length of their cubic box, and they have the same number of dark-matter particles, so they differ in mass resolution by a factor of 2:
\begin{itemize}
\item L86: ~~~86 Mpc, $m_{\rm dm} \! = \! 2.36 \times 10^7 \Msun$
\item L108: 108 Mpc, $m_{\rm dm} \! = \! 4.67 \times 10^7 \Msun$
\item L136: 136 Mpc, $m_{\rm dm} \! = \! 9.32 \times 10^7 \Msun$
\item L172: 172 Mpc, $m_{\rm dm} \! = \! 1.89 \times 10^8 \Msun$
\end{itemize}
Given that \textsc{MUSIC} \citep{HahnAbel2011} allows users to select mass resolution in the zoom-in region in multiples of 8 of the resolution of the box, this suite allows users to generate a zoom-in simulation at any factor of 2 in target mass resolution.

Each simulation stores 11 snapshots, where 0 corresponds to the initial conditions at $z \! = \! 99$.
Each simulation has \textsc{Rockstar} halo catalogs at each snapshot after snapshot 0.
L172 (only) also has halo merger trees from \textsc{Consistent Trees} (see Section~\ref{sec:halo} and the Appendix).

We request users of these cosmological box simulations to cite this article \citep{Wetzel2025}.

\subsection{(Sub)halo catalogs and merger trees}
\label{sec:halo}

Most simulations include a catalog of (sub)halos and their galaxies at each available snapshot, within a directory \texttt{halo/}.
In DR1, this was for the subset of snapshots for which we released the particle data.
DR2 includes (sub)halo catalogs at all available snapshots for most simulations.
Some simulations have no identified halos at the first few snapshots ($z \! \gtrsim \! 15$).
Most simulations have catalogs generated from \textsc{Rockstar} \citep{Behroozi2013a} within \texttt{halo/rockstar\_dm/}, so named because we ran \textsc{Rockstar} using only dark-matter particles.
Some simulations have catalogs generated via the Amiga Halo Finder \citep[\textsc{AHF};][]{Knollmann2009} in \texttt{halo/AHF/}, which uses all particle species.
Some simulations have both types of catalogs.
See \citet{Wetzel2023} for details of these catalogs.

For all \textit{Core} simulations with \textit{Base Physics}, the \textit{Later Reionization} and \textit{Dark Matter Only} suites, and for MW-mass simulations in the \textit{MHD+} and \textit{Cosmic Ray} suites, DR2 includes merger trees across all available snapshots, generated via \textsc{Consistent Trees} \citep{Behroozi2013b} from the \textsc{Rockstar} (sub)halo catalogs.
As with the \textsc{Rockstar} catalogs, we use \href{https://bitbucket.org/awetzel/halo_analysis}{\textsc{HaloAnalysis}}\footnote{\href{https://bitbucket.org/awetzel/halo_analysis}{bitbucket.org/awetzel/halo\_analysis}} to process the output from \textsc{Consistent Trees}, adding additional properties and converting units.
Each simulation stores all trees for all (sub)halos across all snapshots in a single HDF5 file \texttt{tree.hdf5}, within the directory \texttt{halo/rockstar\_dm/catalog\_hdf5/}.
See the Appendix for the contents of \texttt{tree.hdf5}.
For completeness, we also provide the ASCII text file \texttt{halo/rockstar\_dm/catalog/trees/tree\_0\_0\_0.dat} that \textsc{Consistent Trees} directly outputs.
See the documentation from \textsc{Consistent Trees} for the contents of this text file.
The top-level \textit{Core} directory includes a directory \texttt{rockstar\_config/} that contains the file \texttt{consistent\_trees\_config.txt}, which lists the settings we used when running \textsc{Consistent Trees}, alongside the file \texttt{rockstar\_config.txt}, which lists the settings we used when running \textsc{Rockstar}.

\textsc{Consistent Trees} also produces ASCII text files that contain history-based properties for each (sub)halo at each snapshot, named \texttt{hlist\_aaa.list}, where \texttt{aaa} is the expansion scale factor of the snapshot, located within \texttt{halo/rockstar\_dm/catalog/hlists/}.
As \citet{Wetzel2023} describes, each halo catalog file \texttt{halo\_NNN.hdf5} for a simulation with merger trees includes a subset of the history-based properties from \texttt{hlist\_aaa.list}.
Users who want to access the entire contents of \texttt{hlist\_aaa.list} should read that text file, and see the documentation from \textsc{Consistent Trees} regarding such contents.

Almost all (sub)halos in a halo catalog \texttt{halo\_NNN.hdf5} are in the corresponding merger trees \texttt{tree.hdf5}, and almost all (sub)halos at a given snapshot in the merger trees are in the halo catalog.
However, they do not match exactly: some (sub)halos in a catalog do not exist in the trees, and vice versa.
This is for two reasons.
First, \textsc{Consistent Trees} adds ``phantom'' (interpolated) (sub)halos to the merger trees in cases where \textsc{Rockstar} is not able to identify a (sub)halo for one or a few snapshots in a row, for example, when a subhalo orbits close to the center of its host halo.
As the Appendix describes, the merger trees contain a boolean property for each (sub)halo, \texttt{am.phantom}, that indicates if the (sub)halo in the merger trees at that snapshot is a phantom.
Second, \textsc{Consistent Trees} excludes some (sub)halos from the merger trees that it thinks are ``noise'', especially if they are near the resolution limit and/or only exist for one or a few snapshots.

As \citet{Wetzel2023} describes, we use \href{https://bitbucket.org/awetzel/halo_analysis}{\textsc{HaloAnalysis}} to assign star particles (but not gas cells) to \textsc{Rockstar} (sub)halos in post-processing, generating corresponding galaxy stellar properties for each (sub)halo in a file named \texttt{star\_NNN.hdf5}, where \texttt{NNN} is each snapshot index.
We generated \texttt{star\_NNN.hdf5} based on the (sub)halo catalog at each snapshot, not the merger trees.
Thus, the(sub)halos in \texttt{star\_NNN.hdf5} exactly match their order/index in \texttt{halo\_NNN.hdf5}.
If one uses \href{https://bitbucket.org/awetzel/halo_analysis}{\textsc{HaloAnalysis}} to read the halo catalogs, it automatically reads the stellar properties from \texttt{star\_NNN.hdf5} and assigns them to each (sub)halo.
Similarly, if one uses \href{https://bitbucket.org/awetzel/halo_analysis}{\textsc{HaloAnalysis}} to read the merger trees, with a simple input parameter flag one can set it to read the stellar properties from \texttt{star\_NNN.hdf5} and assign them to each (sub)halo in the trees as well.
Alternately, when analyzing (sub)halos in the merger trees, one can use \texttt{catalog.index} to obtain a (sub)halo's index in \texttt{star\_NNN.hdf5}.

\vspace{8 mm}

\section{Accessing data, license, and citing}
\label{sec:access}

The FIRE-2 simulations continue to be available via FlatHUB at:
\href{http://flathub.flatironinstitute.org/fire}{flathub.flatironinstitute.org/fire}.
FlatHUB provides two ways to access the data.
First, using the website above, users can click on ``Browse'' to access the data via a web browser.
We recommend this method to browse the data and download a small amount of it.
Second, the website above provides a Globus ID for transferring via \href{https://app.globus.org}{Globus}.\footnote{
\href{https://app.globus.org}{app.globus.org}
}
We recommend using Globus to transfer a large amount of data.

We release FIRE-2 data under the license \href{https://creativecommons.org/licenses/by/4.0}{Creative Commons BY 4.0}\footnote{\href{https://creativecommons.org/licenses/by/4.0}{creativecommons.org/licenses/by/4.0}}.
We request that anyone using these data to (continue to) cite the published ApJS article that introduced DR1 \citep{Wetzel2023} as the primary citation.
We also welcome citations to this article \citep{Wetzel2025} for DR2 in addition.
Specifically, we request users to cite as follows:

\textit{
We use the publi,cly available FIRE-2 cosmological zoom-in simulations \citep{Wetzel2023, Wetzel2025} from the Feedback in Realistic Environments (FIRE) project, generated using the Gizmo code \citep{Hopkins2015} and the FIRE-2 physics model \citep{Hopkins2018b}.
}

We also request users to cite the individual published article(s) that introduced each simulation used, as listed in Tables 1, 2, 3 in \citet{Wetzel2023} for the simulations first released in DR1, and as we listed in this article for the simulations we release in DR2.
In summary:
\begin{itemize}[leftmargin=*]
\item \textit{Core} simulations with \textit{Base Physics}: cite the work that introduced each one as listed in Table~1 of \citet{Wetzel2023}
\item \textit{Dark Matter Only}: cite the work that introduced the corresponding baryonic simulation as listed in Table~1 of \citet{Wetzel2023}
\item \textit{Later Reionization}: cite \citet{Gandhi2022} for m09; cite this article \citep{Wetzel2025} for m12f, m12i, m12m
\item \textit{MHD+} and \textit{Cosmic Ray}: cite \citet{Hopkins2020b}
\item \textit{Massive Halo}: cite \citet{AnglesAlcazar2017b} for A1, A2, A4, A8; cite \citet{Cochrane2023a} for B1, B2, C1, C2
\item \textit{High Redshift}: cite \citealt{Ma2018a}, \citealt{Ma2019}, \citealt{Ma2020b}, depending on the simulation (see Table~3 of \citealt{Wetzel2023})
\end{itemize}

\section{Acknowledgements}

We generated the FIRE-2 simulations using:
resources at the Texas Advanced Computing Center (TACC), supported by the NSF, including Stampede, Stampede 2, Stampede 3, and Frontera, via the Extreme Science and Engineering Discovery Environment (XSEDE), and the Advanced Cyberinfrastructure Coordination Ecosystem: Services and Support (ACCESS), including allocations TG-AST120025, TG-AST140023, TG-AST140064, TG-AST160048, AST21010 and AST20016;
Blue Waters, supported by the NSF;
Pleiades, via the NASA High-End Computing (HEC) Program through the NASA Advanced Supercomputing (NAS) Division at Ames Research Center, including allocations HEC SMD-16-7592, SMD-16-7561, SMD-17-120;
the Quest computing cluster at Northwestern University;
and the High Performance Computing Core Facility at the University of California, Davis.
This work uses data hosted by the Flatiron Institute's FIRE data hub, and we generated data using the Flatiron Institute's computing clusters \texttt{rusty} and \texttt{popeye}; the Flatiron Institute is supported by the Simons Foundation.
\textit{yt Hub} is supported in part by the Gordon and Betty Moore Foundation's Data-Driven Discovery Initiative through Grant GBMF4561 to Matthew Turk and the National Science Foundation under Grant ACI-1535651.

AW received support from NSF, via CAREER award AST-2045928 and grant AST-2107772.
JS was supported by an NSF Astronomy and Astrophysics Postdoctoral Fellowship under award AST-2102729.
PJG was supported by a Frontera Computational Science Fellowship from TACC while at UC Davis.
DAA acknowledges support from NSF grant AST-2009687 and CAREER award AST-2442788, NASA grant ATP23-0156, STScI JWST grants GO-01712.009-A, AR-04357.001-A, and AR-05366.005-A, an Alfred P. Sloan Research Fellowship, and Cottrell Scholar Award CS-CSA-2023-028 by the Research Corporation for Science Advancement.
RES acknowledges support from NASA grant 19-ATP19-0068; the Research Corporation through the Scialog Fellows program on Time Domain Astronomy; from NSF grant AST-2007232; HST-AR-15809 from STScI.
RF acknowledges financial support from the Swiss National Science Foundation grant PP00P2\_194814.
ZH was supported by a Gary A.~McCue postdoctoral fellowship at UC Irvine.
TKC is supported by the Science and Technology Facilities Council (STFC) astronomy consolidated grant ST/P000541/1 and ST/T000244/1.
LN is supported by the Sloan Fellowship, the NSF CAREER award 2337864, and NSF award 2307788.
SL was supported by NSF grant AST-2109234 and HST-AR-16624 from STScI.
MBK acknowledges support from NSF CAREER award AST-1752913, NSF grants AST-1910346 and AST-2108962, NASA grant NNX17AG29G, and HST grants AR-15006, AR-15809, GO-15658, GO-15901, GO-15902, AR-16159, GO-16226 from STScI.
CAFG was supported by NSF through grants AST-2108230 and AST-2307327; by NASA through grants 21-ATP21-0036 and 23-ATP23-0008; and by STScI through grant JWST-AR-03252.001-A. 
DK was supported by NSF grants AST-1715101 and AST-2108314.
Support for PFH was provided by NSF Research Grants 1911233, 20009234, 2108318, NSF CAREER grant 1455342, NASA grants 80NSSC18K0562, HST-AR-15800.

\bibliographystyle{aasjournal}
\bibliography{biblio}{}

\appendix
\vspace{-6 mm}

\section*{Consistent Trees halo merger trees}
\label{sec:merger_tree}

Here we describe the contents of the file \texttt{tree.hdf5}, which stores the full (sub)halo merger trees for a given simulation.
See \citet{Wetzel2023} for details of the halo/galaxy catalogs.
We first generated (sub)halo catalogs using a slightly \href{https://bitbucket.org/awetzel/rockstar-galaxies}{modified version}\footnote{
\href{https://bitbucket.org/awetzel/rockstar-galaxies}{bitbucket.org/awetzel/rockstar-galaxies}
} of \textsc{Rockstar-Galaxies}\footnote{
\href{https://bitbucket.org/pbehroozi/rockstar-galaxies}{bitbucket.org/pbehroozi/rockstar-galaxies}
} \citep{Behroozi2013a}.
We then generated merger trees using a slightly \href{https://bitbucket.org/awetzel/consistent-trees}{modified version}\footnote{
\href{https://bitbucket.org/awetzel/consistent-trees}{bitbucket.org/awetzel/consistent-trees}
} of \textsc{Consistent Trees}\footnote{
\href{https://bitbucket.org/pbehroozi/consistent-trees}{bitbucket.org/pbehroozi/consistent-trees}
} \citep{Behroozi2013b}.
Finally, we processed these data, adding additional properties and converting units, and rewrote the tree file to HDF5 format via \href{https://bitbucket.org/awetzel/halo_analysis}{\textsc{HaloAnalysis}}\footnote{\href{https://bitbucket.org/awetzel/halo_analysis}{bitbucket.org/awetzel/halo\_analysis}}.

Below, \texttt{tree.index} refers to the (unique) index of a (sub)halo within the full merger trees for a simulation.
All indexing starts at 0.
This is distinct from the \texttt{catalog.index} of each (sub)halo in the halo catalog file at a given snapshot, \texttt{halo\_NNN.hdf5}, where \texttt{NNN} is its snapshot index.
Every (sub)halo in the merger tree file \texttt{tree.hdf5} contains a pointer, \texttt{catalog.index}, to its index in the halo catalog file \texttt{halo\_NNN.hdf5} at its snapshot.
Also, every (sub)halo in each halo catalog file \texttt{halo\_NNN.hdf5} contains a pointer, \texttt{tree.index}, to its index in the merger tree file \texttt{tree.hdf5}.
This makes it easy to link the halo catalogs and merger trees and use them together.

The file named \texttt{tree.hdf5} stores several properties for each (sub)halo.
First, the properties that describe the time evolution (history) of a (sub)halo, which are unique to the merger tree file (not in the halo catalogs), are:
\begin{itemize}
\item \texttt{tid} - tree ID, unique across all (sub)halos across all snapshots; note that all versions of \texttt{index} below refer to a (sub)halo's \texttt{index}, not its \texttt{tid}

\item \texttt{snapshot} - snapshot index

\item \texttt{am.phantom} - whether the (sub)halo is interpolated across snapshots
\item \texttt{am.progenitor.main} - whether the (sub)halo is the main (most-massive) progenitor of its descendant
\item \texttt{progenitor.main.index} - tree index of main (most massive) progenitor
\item \texttt{progenitor.co.index} - tree index of next co-progenitor (with same descendant)
\item \texttt{progenitor.number} - number of progenitors

\item \texttt{descendant.index} - tree index of descendant
\item \texttt{descendant.snapshot} - snapshot index of descendant

\item \texttt{final.index} - tree index at final snapshot
\item \texttt{dindex} - depth-first order (index) within tree
\item \texttt{progenitor.co.dindex} - depth-first index of next co-progenitor (with same descendant)
\item \texttt{progenitor.last.dindex} - depth-first index of last progenitor (earliest in time), including all progenitors
\item \texttt{progenitor.main.last.dindex} - depth-first index of last progenitor (earliest), along main progenitor branch
\item \texttt{major.merger.snapshot} - snapshot index of last major merger

\item \texttt{central.index} - tree index of most-massive central halo; this is always a central (host) halo
\item \texttt{central.local.index} - tree index of the ``local'' (lowest-mass) central; this can be a subhalo of a more massive central halo

\item \texttt{catalog.index} - index (not ID) in the halo catalog at that snapshot \texttt{halo\_NNN.hdf5}
\end{itemize}

Next, the instantaneous properties of (sub)halos that are copied over from the halo catalogs are:
\begin{itemize}
\item \texttt{position} [kpc comoving] - 3-D position, along simulation's x,y,z coordinates
\item \texttt{velocity} [km s$^{-1}$] - 3-D velocity, along simulation's x,y,z coordinates

\item \texttt{mass} or \texttt{mass.200m} [M$_\odot$] - mass of dark matter within $\Rthm$
\item \texttt{mass.bound} [M$_\odot$] - mass of dark matter within $\Rthm$ that is bound to the (sub)halo
\item \texttt{mass.vir} [M$_\odot$] - mass of dark matter within the virial radius defined via \citet{BryanNorman1998}
\item \texttt{mass.200c} [M$_\odot$] - mass of dark matter within $\Rthc$
\item \texttt{mass.lowres} [M$_\odot$] - mass of low-resolution dark matter within $\Rthm$ as computed by \textsc{Rockstar}; this can be inaccurate in some cases, so better to use \texttt{dark2.mass} in \texttt{star\_NNN.hdf5}; we recommend caution regarding any (sub)halo in which the low-resolution mass exceeds a few percent of the total mass
\item \texttt{vel.circ.max} [km s$^{-1}$] - maximum of the circular velocity profile, $\sqrt{G M_{\rm dm}(< r) / r}$
\item \texttt{vel.std} [km s$^{-1}$] - standard deviation of the velocity of member dark-matter particles
\item \texttt{radius} [kpc physical] - $\Rthm$
\item \texttt{scale.radius} [kpc physical] - Navarro–Frenk–White (NFW) scale radius, computed from a fit to the density profile
\item \texttt{scale.radius.klypin} [kpc physical] - NFW scale radius, computed by converting from the radius of \texttt{vel.circ.max} \citep[see][]{Klypin2011}
\item \texttt{axis.b/a} and \texttt{axis.c/a} - ratios of second and third to first largest ellipsoid shape axis \citep{Allgood2006}
\item \texttt{spin.peebles} - spin parameter from \citet{Peebles1969}
\item \texttt{spin.bullock} - spin parameter from \citet{Bullock2001}
\item \texttt{position.offset} [kpc physical] and \texttt{velocity.offset} [km s$^{-1}$] - offset distance and total velocity between the maximum density peak within the halo and the particle average
\end{itemize}

Finally, in the process of creating \texttt{tree.hdf5}, \textsc{HaloAnalysis} also assigns the primary host halo (which hosts the primary galaxy), defined as the most massive halo within the zoom-in region (uncontaminated by low-resolution dark-matter particles).
However, unlike with the halo catalogs, for which \textsc{HaloAnalysis} assigns this primary host independently at each snapshot, for the merger trees \textsc{HaloAnalysis} assigns the primary host at $z \! = \! 0$, and it defines the primary host at all previous snapshots as the main progenitor of the primary host at $z \! = \! 0$.
Hence, the primary host within the halo catalog and the merger trees is almost always the same at $z \! \approx \! 0$, but they can differ at high redshifts, because the main progenitor of the primary host halo at $z \! = \! 0$ is not always the most massive halo at early times \citep[see][]{Santistevan2020}.
Indeed, at sufficiently early times, the classification of a single primary host halo/galaxy progenitor can be ill-defined \citep{Santistevan2020}.

Thus, \texttt{tree.hdf5} also contains the following properties for each (sub)halo, with respect to the center of this primary host halo at the same snapshot:
\begin{itemize}
\item \texttt{host.index} - tree index of the primary host halo at this snapshot
\item \texttt{host.distance} [kpc physical] - 3-D distance, along simulation’s x,y,z coordinates
\item \texttt{host.velocity} [km s$^{-1}$] - 3-D velocity, along simulation’s x,y,z coordinates
\item \texttt{host.velocity.tan} [km s$^{-1}$] - tangential velocity
\item \texttt{host.velocity.rad} [km s$^{-1}$] - radial velocity
\end{itemize}

For the \textit{ELVIS on FIRE} Local Group-like simulations, which contain 2 \ac{MW}-mass host halos, \textsc{HaloAnalysis} also assigns these properties for the second (lower-mass) host, as \texttt{host2.index}, \texttt{host2.distance}, and so on.

\end{document}